\documentclass[aps,floatfix,prd,showpacs]{revtex4}
\usepackage{graphicx}
\usepackage{dcolumn}
\usepackage{bm}

\begin{document}

\def\dq{\frac{d^4q}{(2\pi)^4}\,}
\def\dqE{\frac{d^4q_E}{(2\pi)^4}\,}
\def\be{\begin{equation}}
\def\ee{\end{equation}}

\title{$Z_b$ and $Z_c$ Exotic States as Coupled Channel Cusps}

\author{E.S. Swanson}
\affiliation{
Department of Physics and Astronomy,
University of Pittsburgh,
Pittsburgh, PA 15260,
USA.}

\date{\today}

\begin{abstract}
It is demonstrated that the candidate tetraquark states $Z_b(10610)$, $Z_b(10650)$, $Z_c(3900)$, and $Z_c(4025)$ are coupled channel cusp effects. The model explains in a natural way the masses and quantum numbers of the putative states and the near equality of the widths of the $Z_b$ states. 
It is argued that the $Z_c(3900)$ and $Z_c(4025)$ should be visible in 
$\bar B^0  \to J/\psi \pi^0\pi^0$ or $B^- \to J/\psi \pi^- \pi^0$, but should {\it not} appear in $\bar B^0 \to J/\psi \pi^+\pi^-$, in agreement with recent LHCb results. Additional tests for cusp effects are suggested.
\end{abstract}

\maketitle 

\section{Introduction}

The recent spate of discoveries of exotic heavy mesons has engendered much speculation about their dynamical origins. In particular, the manifestly exotic charged charmonium and bottomonium states have revived old controversies regarding the possible existence of molecular (loosely bound mesonic colour singlets), tetraquark (tightly bound $qq\bar q \bar q$ colour singlets), and diquonium (bound diquarks, $(qq)(\bar q \bar q)$) states and how these may be experimentally distinguished\cite{multiquarkReviews}.

In the light quark sector speculation about multiquark states began more than 40 years ago with a claim that a dynamical scalar isoscalar resonance in $\pi\pi$ scattering is predicted by current algebra, unitarity, and crossing symmetry\cite{brown}. A related idea was proposed by Jaffe, who noted that $qq\bar q \bar q$ states could make up a scalar nonet ($\sigma$, $\kappa$, $f_0(980)$, $a_0(980)$) \cite{jaffe}. This hypothesis has been a rich source of ideas and controversy ever since. Only recently, with the work of Ref. \cite{sigma}, has it been generally accepted that a $\sigma$ resonance even exists. The interpretation of these states, and the existence of the strange analogue state, $\kappa$, remain open issues. In the intervening decades the idea of multiquark states has been applied to a plethora of additional states, $a_0$ and $f_0$ ($K\bar K$)\cite{KK}, $f_1(1420)$ ($K^*\bar K$)\cite{f1}, $f_2(2010)$ ($\phi\phi$)\cite{f2}, and $f_0(1770)$ ($K^*\bar K^*$) \cite{f0}.

The extension of the multiquark angst to the heavy quark sector began in 2004 with the discovery of the $X(3872)$\cite{X}. Its proximity to $D\bar D^*$ threshold and decay  properties have led to the general acceptance that it is a weakly bound system of  $D^0$ and  $\bar D^{0*}$ mesons\cite{Xthy}. Thus the $X(3872)$ is a prototype for molecular states in the heavy quark sector. Subsequent experimental effort has revealed many new resonances, all of which enjoy -- or suffer -- molecular, tetraquark, or diquonium interpretations. Amongst these are the $Z(4475)$\cite{Z4475}, $Z_1(4050)$ and $Z_2(4250)$\cite{Z1}, $Y(4260)$\cite{Y},  $Y(4008)$\cite{Y2},  $G(3900)$\cite{G}, $Y(4140)$\cite{Y3},  and the $Y(4660)$\cite{Y4}.

This paper focusses on four of these states: the manifestly exotic $Z_b(10610)$, $Z_b(10650)$, $Z_c(3900)$, and $Z_c(4025)$. The $Z_b$ states were discovered by Belle\cite{BelleBB} in $e^+e^- \to \Upsilon(5S) \to \Upsilon(nS)\pi^+\pi^-$ and $\Upsilon(5S) \to h_b(nP)\pi^+\pi^-$ via their decays to $\Upsilon(nS)\pi^\pm$ or $h_b(nP)\pi^\pm$. The masses and widths of these resonances were determined to be 
$M= 10608.4 \pm 2.0$ MeV, $\Gamma = 15.6 \pm 2.5$ MeV and 
$M= 10653.2 \pm 1.5$ MeV, $\Gamma = 14.4 \pm 3.2 $ MeV respectively. An examination of angular distributions heavily favours the spin-parity assignment $J^P = 1^+$ for both states.

The $Z_c(3900)$ was discovered by the BESIII collaboration\cite{BESZc} in $e^+e^- \to Y(4260) \to J/\psi \pi^+ \pi^-$ in the charged mode $Z_c \to J/\psi \pi^\pm$. The reported mass and width are $M= 3899.0 \pm 3.6 \pm 4.9$ MeV and $\Gamma = 46 \pm 10 \pm 20$ MeV. The quantum numbers of the state are not known. The $Z_c(4025)$ was observed by BESIII in $e^+e^- \to D^* \bar D^* \pi$ at $\sqrt{s} = 4.26$ GeV\cite{4025-2} and in $e^+e^- \to h_c\pi\pi$ at a variety of energies\cite{4025}. Its mass was determined to be 4026.3(4.5) MeV and 4022.9(2.8) MeV in the respective experiments. The measured widths were 24.8(9.5) MeV and 7.9(3.7) MeV.

The proximity of the $Z_b$ states to $B\bar B^*$ (10604 MeV) and $B^*\bar B^*$ (10650 MeV) threshold has inspired speculation that heavy isovector analogues of the $X(3872)$ state have been discovered\cite{BBmolecules}. Similarly, the $Z_c(3900)$ is close to $D\bar D^*$ threshold at 3879 MeV, while the $Z_c(4025)$ is  close of $D^*\bar D^*$ threshold at 4020 MeV. It is, however, important to observe that these resonances lie {\it above} their respective thresholds, thus the possibility that they are kinematical effects must be considered.

This hypothesis has been explored before by Bugg\cite{Bugg2}, who considered triangle diagrams for the process $\Upsilon(5S) \to \Upsilon\pi\pi$. This diagram proceeds via an $\Upsilon\pi:B^*\bar B^*$ vertex coupled with $B^*:B\pi$ and $\Upsilon: BB^*$ vertices. The loop diagram was then modelled with a simple form that assumed a scale driven by pion exchange. No attempt to fit experimental data was made.  A similar model was constructed by Chen {\it et al.} and applied to the process $Y(4260) \to J/\psi\pi\pi$\cite{chen}. The model amplitude had 10 parameters, which enabled a good fit to the Dalitz plot, thereby explaining the $Z_c(3900)$ as a threshold effect.

The cusp hypothesis is developed further in the following sections where a simple and consistent model that incorporates thresholds is constructed. It is found that all of the relevant experimental results can be explained as cusp effects in a natural fashion with two free parameters. Furthermore, it is argued that the lack of a $Z_c$ signal in the electroweak  decay $\bar B^0 \to J/\psi \pi^+\pi^-$ is a clear indication that the $Z_c$s are indeed an effect due to a coupled channel cusp. The model predicts that both $Z_c$ ``states" may be visible in $\bar B^0 \to J/\psi \pi^0\pi^0$ or $B^- \to J/\psi\pi^-\pi^0$. Similar cusp states are predicted at 10695 MeV and 10745 MeV in $\Upsilon(5S) \to K\bar K \Upsilon(nS)$ due to $B \bar B_s^*$, $B^*\bar B_s$, and $B^* \bar B_s^*$ virtual continuua.

\section{Coupled Channel Cusps}

\begin{figure}[ht]
\includegraphics[width=6cm,angle=0]{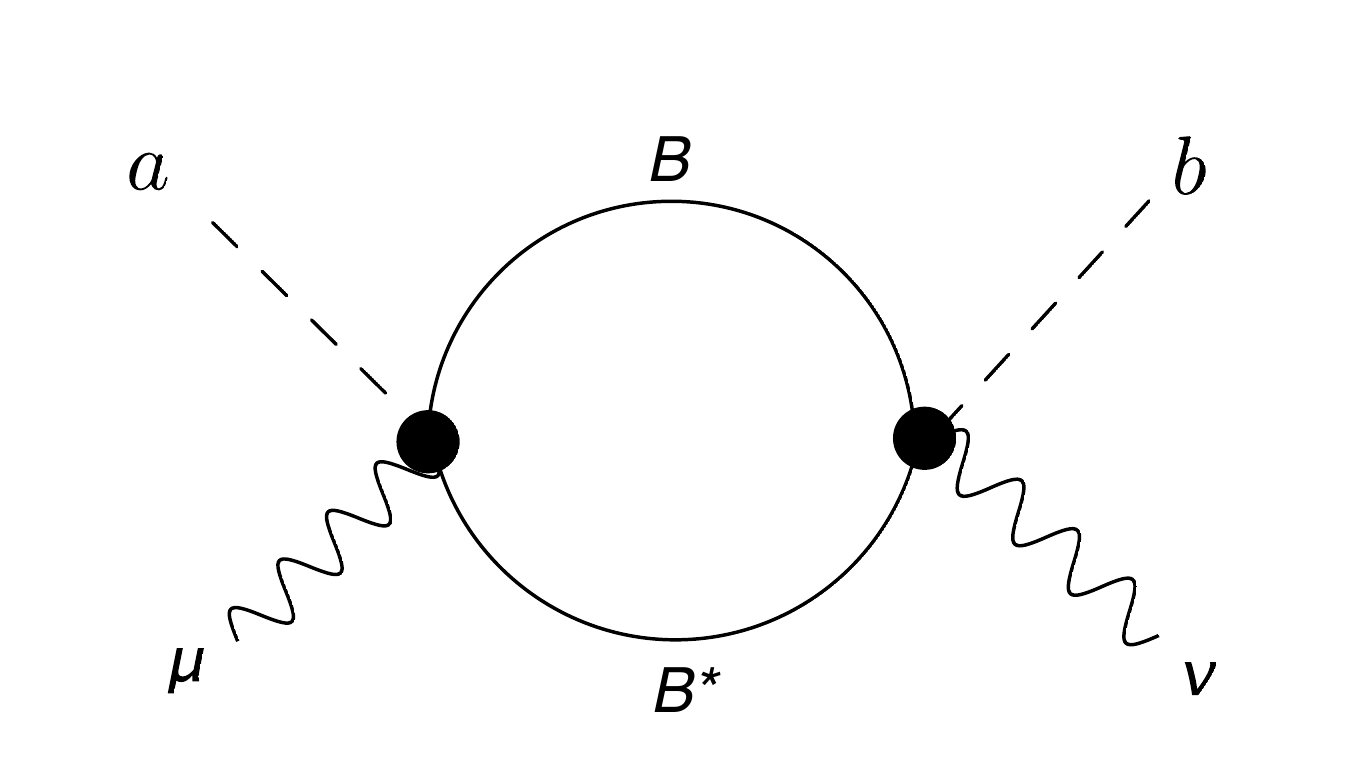}
\caption{Coupled Channels in $\Upsilon\pi$ Scattering.}
\label{fig-cusp}
\end{figure}

The hypothesis is that coupled channel effects can generate signals in Dalitz plots that mimic resonances. A simple model that incorporates this idea can be constructed by considering the crossed channel $\bar \pi \Upsilon \to \pi \Upsilon$ as shown in Fig. \ref{fig-cusp}. Notice that the diagram applies equally well to the reaction $\bar \pi Y(4260)\to \pi J/\psi$ with the replacement of the intermediate particles with their charmed analogues; it also applies to $\bar \pi B \to \pi J/\psi$ if the initial vector particle is replaced with a pseudoscalar. Of course, the diagram should be summed over all intermediate states consistent with the quantum numbers of the reaction.

It is possible to construct an effective field theory to describe this process and evaluate the ensuing diagrams. For example, the vertex shown in the figure can be modelled as

\be
{\cal L} = -i \lambda \Upsilon^\mu \pi^a B^i \frac{\tau^a_{ij}}{2} B^{j*}_\mu.
\ee
However the isospin matrices merely contribute an overall factor, and the momentum dependence induced by spin-dependence in the propagators reduces to a polynomial in $s$. Neither of these effects are central to the physics we pursue, which is the presence of a right hand cut and elastic scattering suppression mediated by the hadronic scale, $\Lambda_{\rm QCD}$.
We therefore eschew the effective field theory approach and simply model the diagram of Fig. \ref{fig-cusp} by writing its imaginary part as

\be
{\rm Im} \Pi_{\alpha\beta}(s) = \sum_i k_i^{1+\ell_{\alpha i}+\ell_{\beta i}} F_{\alpha i}(s) 
F_{\beta i}(s)
\ee
with
\be
k_i^2 = \frac{(s-(m_{1i}+m_{2i})^2)\, (s - (m_{1i}-m_{2i})^2)}{4s}.
\ee
Here $\alpha$ and $\beta$ refer to incoming and outgoing channels, $i$ is a virtual channel consisting of hadrons with masses $m_{1i}$ and $m_{2i}$, and $\ell_{\alpha i}$ is the lowest wave associated with the vertex $\alpha i$, which we assume saturates the given subprocess. The bound state nature of the scattering hadrons is accounted for by a suitably chosen form factor. In the following we shall employ the simple Ansatz

\be
F_{\alpha i} = g_{\alpha i} \exp(-s/2\beta_{\alpha i}^2).
\ee
It is, of course, a simple matter to incorporate nodes or any other structure that is important to the process in question. The scale $\beta_{\alpha i}$ is governed by $\Lambda_{\rm QCD}$.

\begin{figure}[ht]
\includegraphics[width=8cm,angle=0]{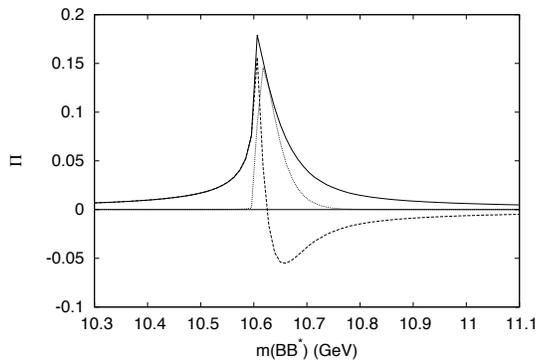}
\caption{Self-energy with $\beta_{BB^*} = 0.7$ GeV. Dashed line: Re $\Pi$; dotted line: Im $\Pi$; solid line: $|\Pi|$.}
\label{fig-pi}
\end{figure}

We now invoke two-body unitarity and the assumption that no resonances contribute to the reaction to obtain the complete analytic ``self energy" portion of the scattering amplitude:

\be
\Pi_{\alpha\beta}(s) = \frac{1}{\pi} \int_{s_{th}}^\infty \, ds' \, \frac{{\rm Im} \Pi_{\alpha\beta}(s')}{s'-s - i\epsilon}
\ee
A typical result for the self energy is shown in Fig. \ref{fig-pi}, which is obtained for the $B\bar B^*$ channel with $\beta_{BB^*} = 0.7$ GeV.

Under these assumptions the spin-averaged scattering amplitude is proportional to the self-energy,

\be
 \overline{|{\cal M}_{\alpha\beta}|^2} \propto |\Pi_{\alpha\beta}(s)|^2.
\ee
The decay amplitude is given by the scattering amplitude with the initial pion momentum reversed and symmetry between the outgoing pions accounted for. The simple bubble structure of the model permits unitarisation if this is deemed necessary. In this case we expect the hadronic rescattering amplitudes to be small so that unitarisation is not important. It is also not relevant to demonstrating the effects of coupled channel cusps and form factors.

Finally, for this mechanism to be relevant it is necessary to demonstrate that the reaction proceeds via the loop in question. This demonstration, of course, depends on the reaction under consideration, hence it will be deferred to the next section where different processes are considered in turn.

\section{Comparison with Experiment}

Since the purpose of this computation is not a detailed analysis  of experimental results, a fit to the Dalitz plot densities will not be attempted. Any fit would also depend sensitively on $\pi\pi$ dynamics in the Dalitz plot, which is not of concern here. Thus Figs. \ref{fig-belle} -- \ref{fig-bess2} present direct computations of the $\Upsilon(nS)\pi$ invariant mass distribution and plot these with relevant experimental data.

Under these conditions the parameters available to the model are the scales $\beta_{\Upsilon(5S)\pi:B\bar B^*}$, $\beta_{\Upsilon(5S)\pi:B^*\bar B^*}$, $\beta_{\Upsilon(nS)\pi:B\bar B^*}$, $\beta_{\Upsilon(nS):B^*\bar B^*}$ and the couplings $g_{\Upsilon(5S)\pi:B\bar B^*}$, $g_{\Upsilon(5S)\pi:B^*\bar B^*}$, $g_{\Upsilon(nS)\pi:B\bar B^*}$, $g_{\Upsilon(nS):B^*\bar B^*}$. Analgous quantities apply to the charmed sector. In the following we shall set

\be
g^2_{nBB^*} = g_{\Upsilon(5S)\pi:B\bar B^*} \cdot g_{\Upsilon(nS)\pi: B\bar B^*}
\ee
with a similar expression for $g^2_{nB*B*}$.

Although this parameter set is dramatically smaller than those typically used to fit the Dalitz plot distribution, it is still too extensive for our purposes. We therefore make drastic, but reasonable, further assumptions: $\beta_{\alpha i} = \beta = 0.7$ GeV for {\it all} channels and $g^2_{nBB*}$ = 0.9 $\cdot$ $g^2_{nB^*B*}$, for all $n$. The latter relationship simply means that the $B\bar B^*$ channel is slightly reduced in strength compared to the $B^*\bar B^*$ channel. These parameters were obtained with a rough fit to the $\Upsilon(3S)\pi$ invariant mass distribution of the $\Upsilon(5S)\to \Upsilon(3S)\pi$ Dalitz plot, as shown below, and will be called the ``canonical fit" in the following. Finally, the remaining free product of couplings is fixed by normalising to the experimental data. Notice that if the model is accurate it should be possible to fit results obtained from a single experiment, like $\Upsilon(5S) \to \Upsilon(nS)\pi\pi$ for $n=1, 2, 3$, with a {\it single} normalization. This provides an important test of the formalism.

\subsection{$\Upsilon(5S) \to \Upsilon(nS)\pi\pi$}

We now seek to compare the predictions of the cusp model with the Belle data for $\Upsilon(5S) \to \Upsilon(nS) \pi\pi$, with $n$ = 1, 2, 3\cite{BelleBB}. The first issue to address is the relative importance of the $B\bar B^*$ loop diagram to the decay process. In this  case we can turn to direct measurements of the couplings of the $\Upsilon(5S)$ to various final states\cite{Up}. It is thereby learned that the $\Upsilon(5S)$ decays to states with no open bottom a scant 3.8\% of the time. Alternatively, decays to $B^{(*)}\bar B^{(*)}$  amount to 57.3\% of the $\Upsilon(5S)$ width, while those to $B^{(*)}\bar B^{(*)}\pi$ account for an additional 8.3\%. Direct coupling to $\Upsilon(nS)\pi\pi$ is always less than $7.8\cdot 10^{-3}$. 

It appears that the dominant diagram is one in which the $\Upsilon(5S)$ fluctuates into a $B^{(*)}\bar B^{(*)}$ pair which then couples to $\Upsilon(nS)\pi\pi$. This represents a bubble in the $s= M_{\Upsilon(5S)}^2$ channel, which thus contributes an approximately constant background. 
It therefore appears likely that the most important diagrams with structure are those coupling to $B^{(*)}\bar B^{(*)}\pi$, as postulated in this approach.

\begin{figure}[ht]
\includegraphics[width=8cm,angle=0]{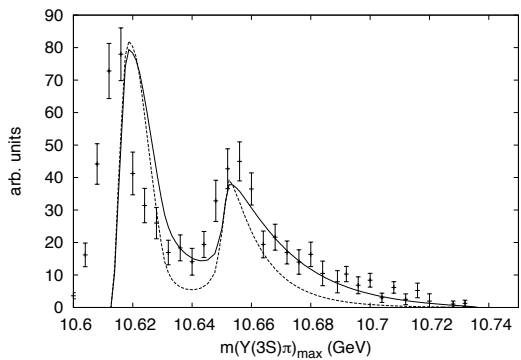}
\caption{\protect Cusp Effects in $\Upsilon(5S) \to \Upsilon(3S)\pi\pi$ and Belle data\cite{BelleBB}. Solid line: $g^2_{3BB^*} = 0.9\, g^2_{3B^*B^*}$, $\beta=0.7$ GeV. Dashed line: $g^2_{3BB^*} = g^2_{3B^*B^*}$, $\beta = 0.5$ GeV.}
\label{fig-belle}
\end{figure}

Our first comparison will be to the reaction $\Upsilon(5S) \to \Upsilon(3S)\pi\pi$ because this process displays little interference from pion dynamics, and thus provides a relatively clean starting point for making comparisons.
Fig. \ref{fig-belle} presents the Belle data for this process along with the model prediction for the case $g^2_{3BB*} = g^2_{3B^*B^*}$, $\beta = 0.5$ GeV (dashed line). Although this model is certainly over-simplified, the agreement with the data is satisfactory. Increasing $\beta$ to 0.7 GeV and reducing the relative size of the $B\bar B^*$ cusp by 10\% yields the solid curve in the figure. This parameter set was called the ``canonical fit" above. This very simple description of the data should be contrasted to that obtained in a resonance interpretation, where relative phases, strengths, pole locations, and pole widths are all completely free parameters.

\begin{figure}[ht]
\includegraphics[width=8cm,angle=0]{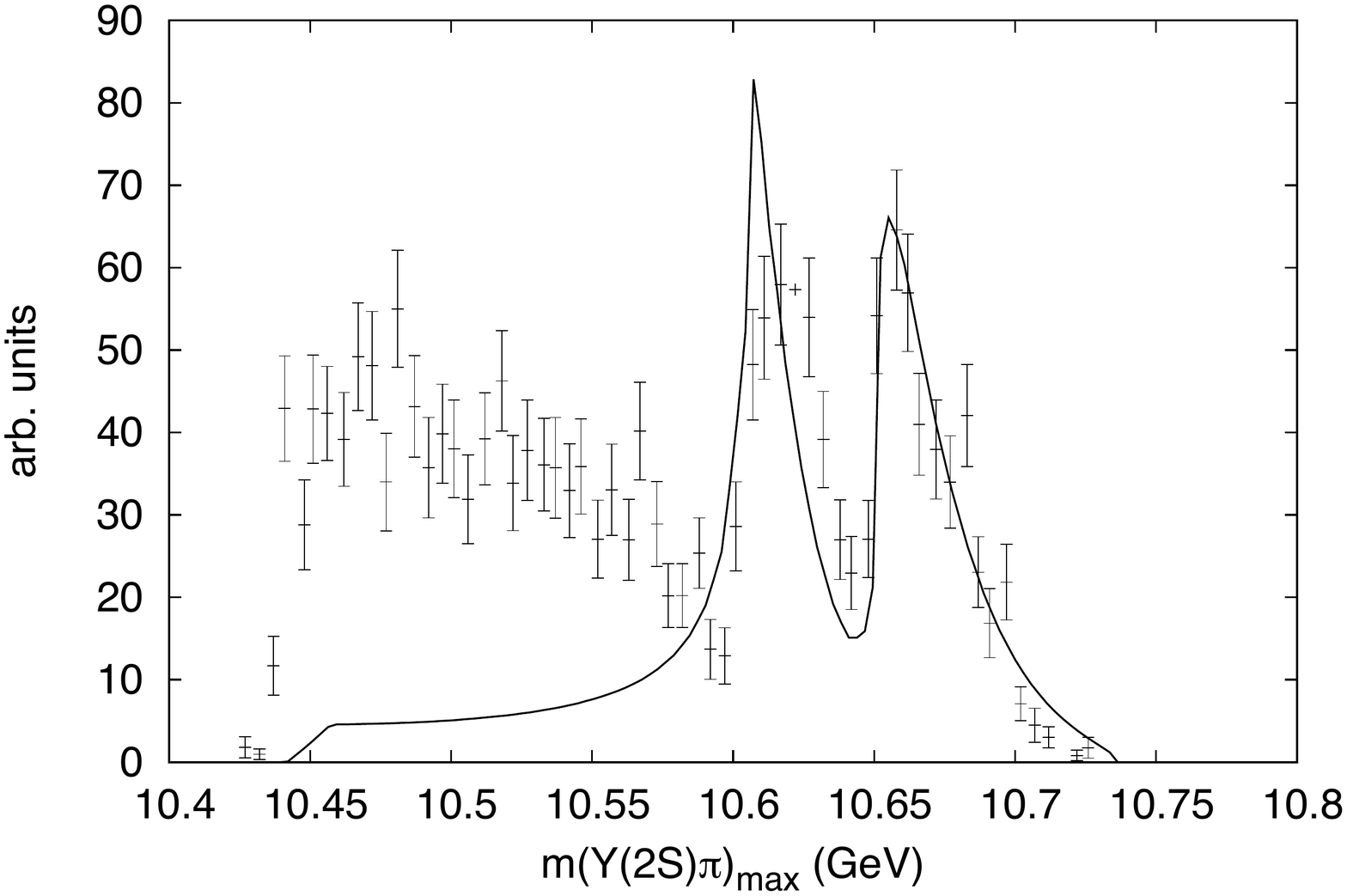}
\includegraphics[width=8cm,angle=0]{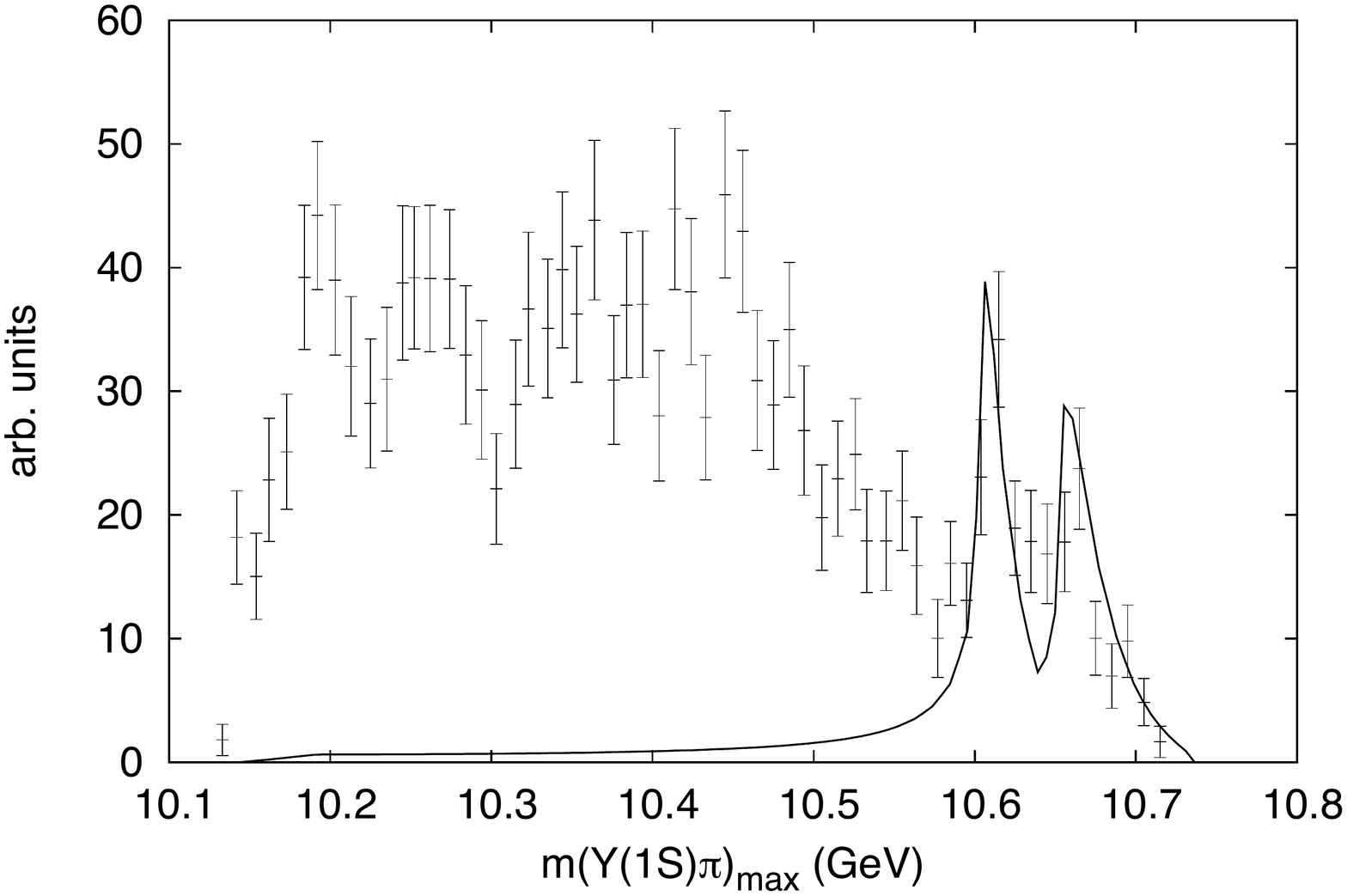}
\caption{\protect Cusp Effects in $\Upsilon(5S) \to \Upsilon(2S)\pi\pi$ (left),  $\Upsilon(5S) \to \Upsilon(1S)\pi\pi$ (right), and Belle data\cite{BelleBB}. Solid line: canonical fit. Many $\pi\pi$ resonances contribute to the structure seen at low invariant mass.}
\label{fig-belle2}
\end{figure}

With the canonical fit in place, the comparison to the $\Upsilon(2S)$ and $\Upsilon(1S)$ data can be made. These are presented in Fig. \ref{fig-belle2}, where it is seen that both sets of ``resonance" peaks are reproduced very well. Again, no parameters have been adjusted, in stark contrast to a resonance fit, which would vary relative Breit-Wigner  strengths and phases for each of these (assuming that the pole positions were fixed by the $\Upsilon(3S)$ data). Furthermore, the normalization used to obtain the $\Upsilon(2S)$ was {\it the same} as that used for the $\Upsilon(3S)$ data, indicating that the simple guess $g^2_{2BB^*} = g^2_{3BB^*}$ and $g^2_{2B^*B^*} = g^2_{3B^*B^*}$ is correct.  This stunning success is only approximately reproduced in the case of the $\Upsilon(1S)$, where we found $g^2_{1BB} \approx 0.7 g^2_{3BB}$ (here $BB$ refers to $B\bar B^*$ and $B^*\bar B^*$).

Because the lowest partial waves in the channels used in the model are $S$-waves, the model predicts that the quantum numbers of the erstwhile resonances are $J^P=1^+$, in agreement with the angular analysis of Belle\cite{BelleBB}. Furthermore, the cusp model naturally predicts that the widths of the ``resonance" states should be in the ratio of $\beta_{\Upsilon(nS)\pi:B\bar B^*}$ to $\beta_{\Upsilon(nS)\pi:B^*\bar B^*}$, which should be approximately unity. Indeed, the measured widths are 15.6 MeV and 14.4 MeV respectively. Again, there is no natural reason for this coincidence to occur in a resonance model.  Finally, there is no reason for the relative phases of the $B\bar B^*$ and $B^*\bar B^*$ channels to differ between the $\Upsilon(nS)\pi$ final states, or for this phase to differ from zero. In fact Belle quote relative phases of $53 \pm 61^{+5}_{-50}$ degrees (1S), $-20 \pm 18^{+14}_{-9}$ degrees (2S), and $6 \pm 24^{+23}_{-59}$ degrees (3S) -- in agreement with model expectations\cite{BelleBB}.

\subsection{$\Upsilon(5S) \to h_b(nP)\pi\pi$}

\begin{figure}[ht]
\includegraphics[width=8cm,angle=0]{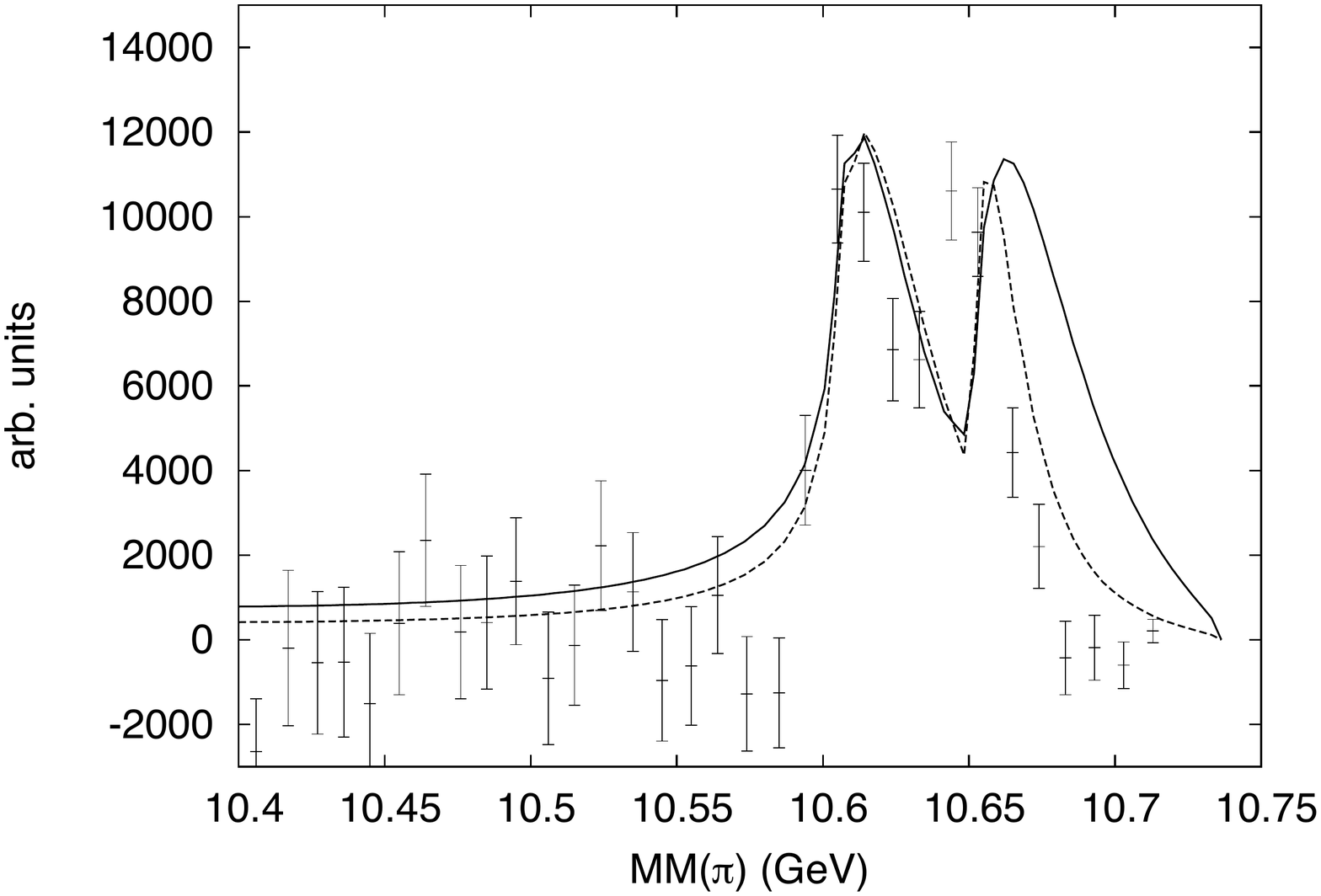}
\includegraphics[width=8cm,angle=0]{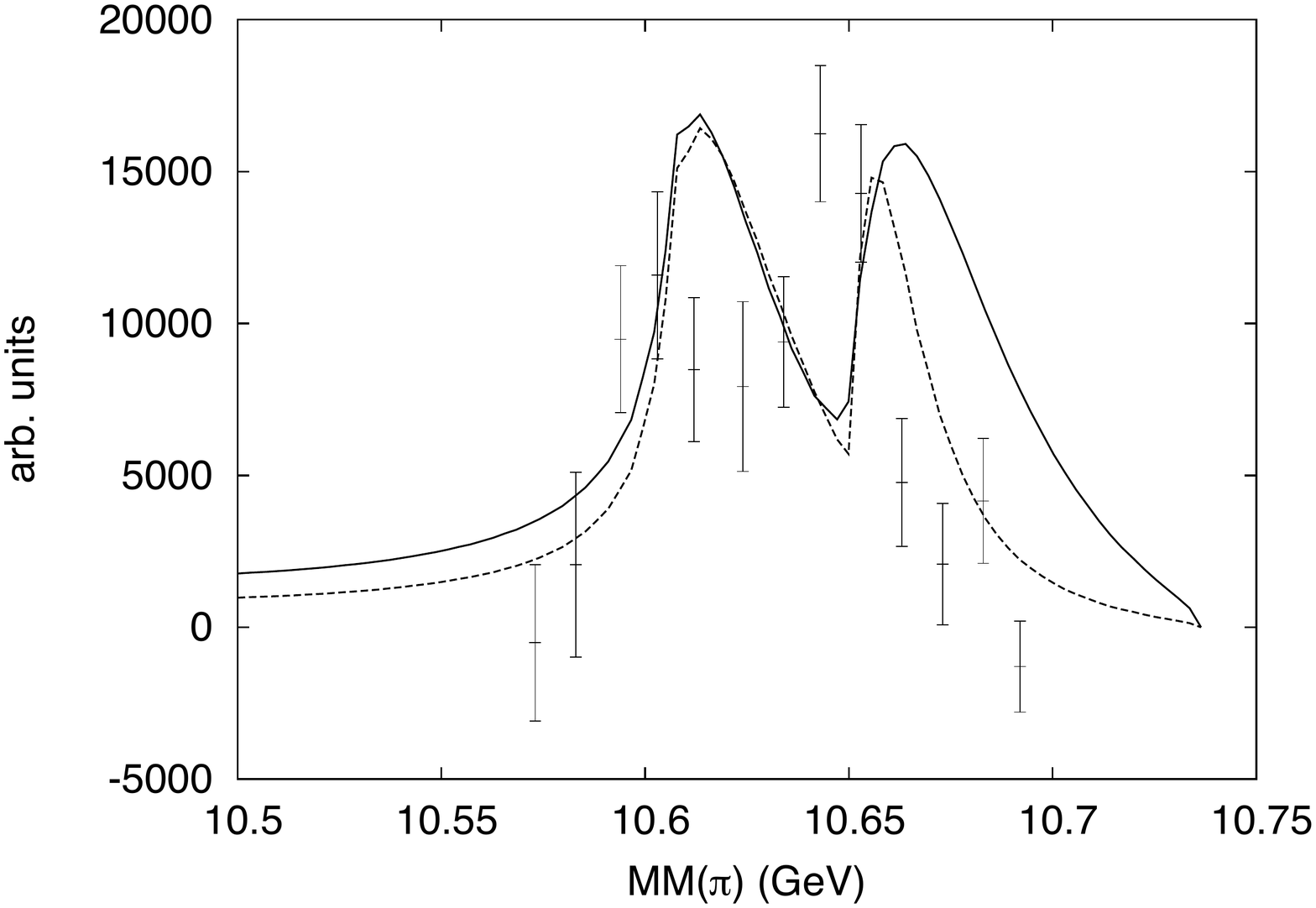}
\caption{\protect Cusp Effects in $\Upsilon(5S) \to h_b(1P)\pi\pi$ (left), $\Upsilon(5S) \to h_b(2P)\pi\pi$ (right), and Belle data. Solid line: canonical fit. Dashed line: $\beta_{B\bar B^*} = 0.7$ GeV, $\beta_{B^*\bar B^*} = 0.4$ GeV, $g^2_{B\bar B^*} = 0.5 g^2_{B^*\bar B^*}$.}
\label{fig-hb}
\end{figure}

The Belle collaboration also measured distributions for $\Upsilon(5S)\to h_b(nP)\pi\pi$\cite{BelleBB}. In this case one has  $\ell_{\Upsilon(5S)\pi:BB}+ \ell_{h_b\pi:BB} = 1$ rather than 0 as in the $\Upsilon(nS)$ cases. The comparison to data is shown in Fig. \ref{fig-hb}. The canonical fit matches the data quite well, although in this case some further simple parameter adjustments can improve the description. The dashed line indicates one such modification, wherein the scale of the $h_b\pi:B^*\bar B^*$ form factor was reduced to 0.4 GeV and the relative strengths between the cusps was lowered to 0.5.

\subsection{$Y(4260) \to J/\psi \pi\pi$}

The BESIII collaboration has seen a state analogous to the $Z_b(10610)$ in $Y(4260)$ decays to $J/\psi \pi\pi$\cite{BESZc}. The model employed so far can also account for this process by mapping bottom to charm quarks, $\Upsilon(5S)$ to the $Y(4260)$, and $\Upsilon(nS)$ to $J/\psi$. Unfortunately, the decay modes of the $Y(4260)$ are largely  unknown and one cannot infer that $D^{(*)}\bar D^{(*)}$ loops dominate its decay. However, the decay patterns of the $\psi(4415)$, $\psi(4160)$, and the $\psi(4040)$ mirror those of the $\Upsilon(5S)$ with respect to open and hidden flavour channels and there is no reason to assume otherwise with the $Y(4260)$. 

\begin{figure}[ht]
\includegraphics[width=8cm,angle=0]{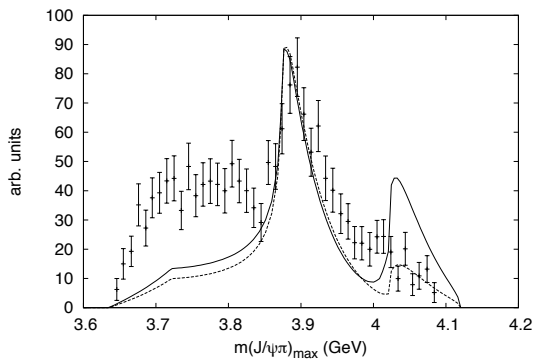}
\caption{\protect Cusp Effects in $Y(4260) \to J/\psi \pi\pi$ and BESIII data\cite{BESZc}. Solid line: canonical fit; dashed line: $g^2_{BB^*} = 2 g^2_{B^*B^*}$, $\beta = 0.7$ GeV.}
\label{fig-bess}
\end{figure}

A comparison of the model to the BESIII data is displayed in Fig. \ref{fig-bess}. The peak at $Z_c(3900)$ is reproduced well. Notice, however, that the cusp model predicts a ``resonance" at 4025 MeV that should be visible in the data. As with the $Z_b$ states, its width should match its sister state's, which should therefore be approximately 40 MeV. 
While there are hints of a bump in the BESIII data near the edge of phase space, this does not match the prediction of the canonical fit very well. Of  course, the canonical parameters were heavily restricted and perhaps some variation in model parameters should be allowed. Simply changing the ratio of the cusp amplitudes drops the $Z_c(4025)$ peak as shown by the dashed line in the figure. In view of this, it is of interest to search for the $Z_c(4025)$ at higher $\sqrt{s}$ where it will be more apparent in the Dalitz plot. In fact, this  has been done by the BESIII collaboration, as seen in the next section.

\subsection{$e^+e^- \to \pi^+\pi^-h_c$}

\begin{figure}[ht]
\includegraphics[width=8cm,angle=0]{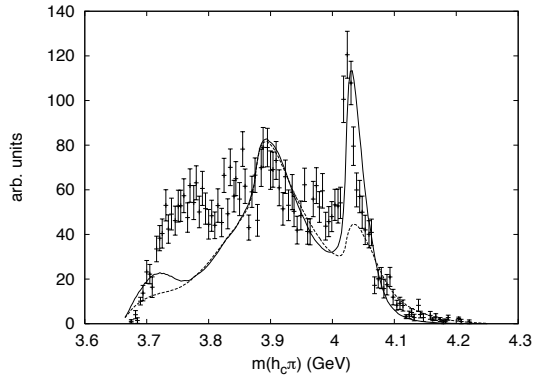}
\caption{\protect Cusp Effects in $e^+e^- \to h_c\pi\pi$ and BESIII data\cite{4025}. Dashed line: canonical fit; solid line: $g^2_{D\bar D^*} = 0.15\, g^2_{D^*\bar D^*}$, $\beta_{DD^*} = 0.7$ GeV, $\beta_{D^*D^*} = 0.4$ GeV.}
\label{fig-bess2}
\end{figure}

The BESIII collaboration has claimed the discovery of a heavy $Z_c$ resonance with a mass near 4025 MeV and a narrow width of 8-25 MeV\cite{4025-2,4025}. The discovery mode was an energy scan with 13 values of $\sqrt{s}$ from 3.90 to 4.42 GeV. Events with a final state of $h_c\pi\pi$ were selected and the sum of these events over all energies revealed a narrow structure at 4025 MeV (see Fig. \ref{fig-bess2}).

These data were modelled by generating 13 Dalitz plots corresponding to the experimental values of $\sqrt{s}$ and summing these with a weight given by the reconstructed number of $h_c$ mesons for each energy. The result with canonical parameters is shown as a dashed line in Fig. \ref{fig-bess2} where one sees a reasonable, but not a good, reproduction of the data. This situation is sufficiently different from the proceeding that some parameter variation is perhaps allowed. The solid line shows the fit with $\beta_{D^*\bar D^*} = 0.4$ GeV, $\beta_{D\bar D^*} = 0.7$ GeV, and $g^2_{D\bar D^*} = 0.15\, g^2_{D^*\bar D^*}$. Notice that these form factor scales are the same as were preferred in $\Upsilon(5S) \to h_b \pi\pi$, indicating an intriguing similarity between charm and bottom systems.

BESIII also observe the $Z_c(4025)$ in $e^+ e^- \to (D^*\bar D^*)^\mp \pi^\pm$ at $\sqrt{s} = 4.26$ GeV \cite{4025-2}. In the cusp model this process occurs via the left hand vertex of Fig. \ref{fig-cusp} and is explained as a threshold enhancement. Threshold enhancements occur whenever a channel opens and the phase space which grows like $p_f^{2\ell+1}$ is moderated by a form factor that decreases rapidly with scale $\Lambda_{QCD}$; these are therefore generic features of any hadronic interaction.

\subsection{Cusps in $B$ Decays}

An important element of the case for cusp effects in $\Upsilon$ decay and $e^+e^-$ processes is that coupling to open flavour channels is preferred. The argument for this has already been made for $\Upsilon$ decay. In the case of $e^+e^-$, open-flavour dominance can be understood as a consequence of hadronisation wherein charm and anticharm (or bottom and antibottom) quarks have opposite and large momenta in the parent rest frame. The resulting flux tube rapidly creates a light quark pair and the system evolves into $D$ mesons moving back-to-back. If these $D$  mesons have high relative momentum they will leave the interaction region and little final state rescattering will occur. If they have low relative momentum (such as for $Y(4260)$ decay) rescattering can be important.

\begin{figure}[ht]
\includegraphics[width=12cm,angle=0]{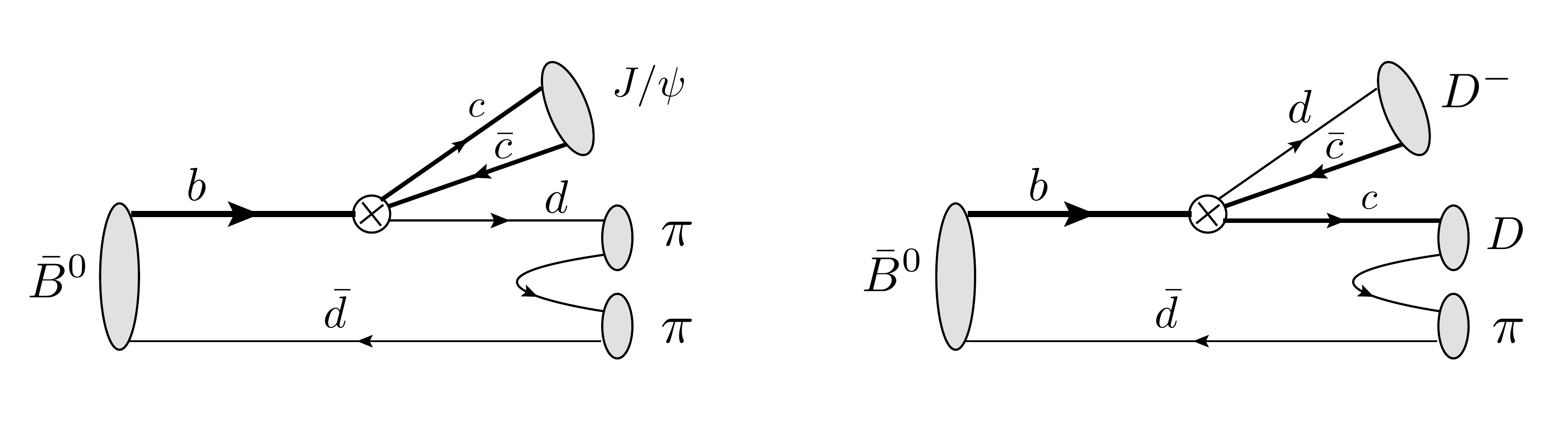}
\caption{Direct process (left); colour-enhanced indirect process I (right).}
\label{fig-decays-1}

\end{figure}

\begin{figure}[ht]
\includegraphics[width=12cm,angle=0]{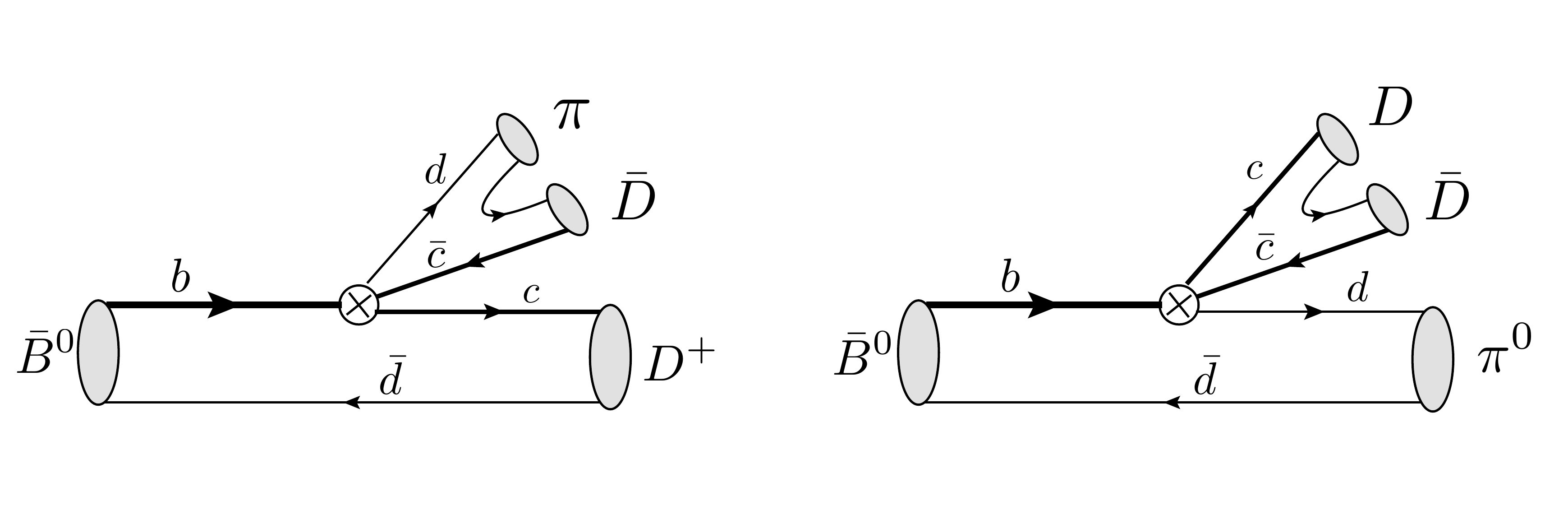}
\caption{Colour-enhanced indirect process II (left); wavefunction-enhanced indirect process (right).}
\label{fig-decays-2}
\end{figure}

An intricate interplay of all of these effects occurs in electroweak $B$ decays, which thereby provide an intriguing entree into the physics of cusp effects. Consider the process $B \to J/\psi \pi\pi$: in principle, a $Z_c(3900)$ and $Z_c(4025)$ should be visible in the final state $J/\psi\pi$ invariant mass distribution. In more detail, the amplitudes that contribute to $\bar B_0\to J/\psi \pi^+\pi^-$ are the colour-suppressed decay wherein the produced charm and anti-charm quarks form a $J/\psi$, which is called the ``direct" amplitude here (Fig. \ref{fig-decays-1}, left panel), and indirect amplitudes that form $D$ mesons and must rescatter to make the final state. Three main topologies for indirect amplitudes are shown in Fig. \ref{fig-decays-1} and Fig. \ref{fig-decays-2}. The first two occur via a colour-enhanced electroweak process with subsequent hadronisation, while the third (right panel Fig. \ref{fig-decays-2}) hadronises the colour-suppressed process. The rescattering required to produce $J/\psi\pi\pi$ inhibits all of the indirect amplitudes, but this is countered by an enhancement with respect to the direct amplitude. In the colour-enhanced case, this is simply the factor $N_c$. Note, however, that the $D$ meson momenta are dominantly opposite in the $B$ meson rest frame, and therefore the $D$ mesons have a suppressed final state interaction.

Alternatively, in the wavefunction-enhanced process the $D\bar D$ system recoils against the pion and final state rescattering can be strong. 
Although the rescattering diagram is suppressed by the final state interactions it is also enhanced with respect to the direct production diagram because it is difficult for the outgoing charm quarks to form a $J/\psi$ meson. Indeed, if one ignored the $d$ quark in the transition $b \to c \bar c d$, the charm quarks would have momenta of about 2 GeV in the $b$ rest frame. An explicit calculation with a constituent quark model then gives a $J/\psi$ formation probability of less than 1\%. 
A more detailed calculation can be made by evaluating the average charm quark momentum in the $c\bar c d$ Dalitz plot. The result in the $c\bar c$ rest frame is $\langle p_c\rangle = 0.92$ GeV.  This implies a capture probability of approximately 25\%, which means that the rate for the direct diagram is suppressed by a factor of 16 with respect to the rescattering diagram. This estimate can be confirmed by noting that typical branching fractions for $B\to X\pi$ are comparable to those for $B\to X\pi\pi$. Furthermore, $BF(B\to D^{(*)}\bar D^{(*)}) \approx 10^{-3}$ whereas $BF(B \to J/\psi\pi\pi) \approx 10^{-5}$. 
Thus it appears that the indirect wavefunction-enhanced channel is an order of magnitude or more larger than the direct channel for reactions in which it contributes.

The electroweak decay $\bar B^0 \to J/\psi \pi\pi$ has recently been measured by the LHCb collaboration\cite{lhc}. While the main point of this research was an attempted analysis of the structure of the $\sigma$ and $f_0(980)$ light mesons, a distribution of events in $J/\psi \pi$ invariant mass was also published. A comparison of the analogous distribution from BESIII reveals a stark difference: although the distributions are over nearly identical mass ranges, there is no sign of the $Z_c(3900)$ or $Z_c(4025)$ in the LHCb data.  This is difficult to understand, because, other than quantum numbers, there is little difference between $\gamma^* \to c\bar c$ ($Y(4260)$ decay where the $Z_c(3900)$ is seen) and $b \to c\bar c d$ ($B$ decay, where the $Z_c$ is not seen).


The process $\bar B_0 \to J/\psi \pi^+\pi^-$ can occur via the direct and colour-enhanced indirect mechanisms. Since no $Z_c$ is observed, one must conclude that  colour-enhanced rescattering is weak and hence no cusp effects are visible. This conclusion is confirmed by the recent measurement of $\bar B_s \to J/\psi K^+K^-$ by LHCb\cite{lhcb-KK}, which finds no evidence of a $J/\psi K^\pm$ resonance. If loops dominated this process cusp ``states" should be visible at 3980 MeV ($D_s \bar D^*$ + $D\bar D_s^*$) and 4125 MeV ($D_s^*\bar D^*$).

It appears that colour-enhanced indirect processes do not give rise to cusp effects. The remaining possibility is wavefunction-enhanced rescattering, which contributes to reactions such as $\bar B_0 \to J/\psi\pi^0\pi^0$, $B^\pm \to J/\psi \pi^\pm \pi^0$, and  $\bar B_s \to J/\psi \pi \varphi$. It is thus of interest to examine these reactions to gain insight into the nonperturbative effects being considered here, and QCD in general.

\section{Conclusions}

It has been argued that the appearance of open flavour thresholds in intermediate states is sufficient to generate signals that mimic resonances that are consistent with the $Z_b(10610)$, $Z_b(10650)$, $Z_c(3900)$, and $Z_c(4025)$.  Model assumptions are that lowest partial waves dominate a given amplitude and that a simple form factor controlled by hadronic scales is sufficient to describe relevant subprocesses. Further simplifications that are not necessary, but add strength to the conclusions, were that a universal scale describes all form factors (except when $h_b$ or $h_c$ mesons are in the final state) and that all couplings are approximately equal. The resulting description of 13 peaks in 7 invariant mass distributions was sufficiently good to imply that the cusp model is a parsimonious and accurate representation of the modelled physics.

The coupled channel cusp model makes several predictions:

(i) $Z$ resonances are $1^+$ states;

(ii) $Z$ resonances lie slightly above open flavour thresholds;

(iii) threshold partners have approximately the same width if they are observed in the same channel; unlike T-matrix poles, this  width can differ in different channels;

(iv) $Z_c$ ``states" may appear in $\bar B^0 \to J/\psi \pi^0 \pi^0$ 
and $B^\pm \to J/\psi \pi^\pm\pi^0$;

(v) similarly, $\bar B_s \to J/\psi \varphi \varphi$ and  $\bar B_0 \to J/\psi \varphi K$ should exhibit cusp effects at $D_s \bar D_s^*$ and $D_s^*\bar D_s^*$ thresholds, while $\bar B_0 \to J/\psi \eta K$ will display $D\bar D^*$, $D^*\bar D^*$, $D_s\bar D_s^*$, and $D_s^*\bar D_s^*$ cusp enhancements;

(vi) it should be possible to discern a rich spectrum of exotic states at higher centre of mass energy in $\Upsilon\pi\pi$. These include a $D_0\bar D_1$ state at 4740 MeV and $D_2\bar D_1$ enhancement at 4880 MeV;

(vii) $\Upsilon(5S) \to K\bar K \Upsilon(nS)$ should show enhancements at 10695 MeV ($B \bar B_s^*$ and $B^*\bar B_s$) and 10745 MeV ($B^*\bar B_s^*$).

Most of these predictions are unnatural in tetraquark or molecular models (except point (i) for molecules).  Point (ii) is in direct conflict with molecular models unless unusual dynamics are postulated.

The molecular candidate $X(3872)$ raises interesting questions in light of the cusp mechanism. The fact that its binding energy and width appear to be below 1 MeV\cite{Xnew} indicate that purely kinematical effects as advocated here are not wholly responsible for this signal. It will be interesting to study the interplay of kinematics and dynamics for this state.

Finally, the relatively large rate for the reaction $\Upsilon(5S) \to h_c\pi\pi$ is somewhat mysterious since a heavy quark spin flip is required to make the transition from a ${}^{3}S_1$ to a ${}^1P_1$ bottomonium state. It is tempting to speculate that the spin flip is being facilitated by the presence of light quark degrees of freedom in the intermediate state that persist over long time scales. In effect, the virtual $B^{(*)}\bar B^{(*)}$ permit the pions to carry off the spin component necessary to affect the $b$ spin flip.

The evidence presented here makes it very likely that the $Z_b$ and $Z_c$ states are kinematical artefacts. The general lesson is that the interpretation of bumps is necessarily dependent on model assumptions. It is therefore important that experimental collaborations examine cusp effects when fitting data where channel thresholds are known to exist.

\acknowledgments

This research was supported in part by the U.S. Department of Energy under contract DE-FG02-00ER41135.

\end{document}